\documentclass[twocolumn,preprintnumbers,amsmath,amssymb]{revtex4}

\usepackage{times}
\usepackage{epsfig}
\usepackage{graphicx}
\usepackage{amsmath}
\usepackage{amssymb}
\usepackage{nicefrac}
\usepackage[caption=false,lofdepth,lotdepth]{subfig}

\newcommand{\Itemize}{\begin{itemize}\setlength{\itemsep}{0pt}}

\def\eg		{{\it e.g.,~}}

\def\plotfive#1#2#3#4#5{\centering \leavevmode
    \epsfxsize=.192\textwidth \epsfbox{#1}
    \hfil \epsfxsize=.192\textwidth \epsfbox{#2}
    \hfil \epsfxsize=.192\textwidth \epsfbox{#3}
    \hfil \epsfxsize=.192\textwidth \epsfbox{#4}
    \hfil \epsfxsize=.192\textwidth \epsfbox{#5}}
\def\plotsix#1#2#3#4#5#6{\centering \leavevmode
    \epsfxsize=.162\textwidth \epsfbox{#1}
    \hfil \epsfxsize=.162\textwidth \epsfbox{#2}
    \hfil \epsfxsize=.162\textwidth \epsfbox{#3}
    \hfil \epsfxsize=.162\textwidth \epsfbox{#4}
    \hfil \epsfxsize=.162\textwidth \epsfbox{#5}
    \hfil \epsfxsize=.162\textwidth \epsfbox{#6}}

\begin{document}

\preprint{APS/123-QED}

\title{Scalable Automated Detection of Spiral Galaxy Arm Segments}

\author{Darren R. Davis, Wayne B. Hayes\\
University of California, Irvine\\
Irvine, California 92697-3435 USA\\
{\tt\small drdavis@ics.uci.edu, wayne@ics.uci.edu}
}

\begin{abstract}

Given an approximately centered image of a spiral galaxy, we describe
an entirely automated method that finds, centers, and sizes the galaxy
and then automatically extracts structural information about the spiral arms.
For each arm segment found, we list the pixels in that
segment and perform a least-squares fit of a logarithmic spiral arc to
the pixels in the segment.
The algorithm takes about 1 minute per galaxy, and can easily be scaled
using parallelism. We have run it on all $\sim$644,000 Sloan objects
classified as ``galaxy'' and large enough to observe some structure. Our
algorithm is stable in the sense that the statistics across
a large sample of galaxies vary smoothly based on algorithmic parameters,
although results for individual galaxies can sometimes vary in a non-smooth
but easily understood manner.
We find a very good correlation between our quantitative
description of spiral structure and the qualitative description provided
by humans via Galaxy Zoo.
In addition, we find that pitch angle often varies significantly
segment-to-segment in a single spiral galaxy, making it difficult to
define ``the'' pitch angle for a single galaxy.
Finally, we point out how complex arm structure (even of long arms) can
lead to ambiguity in defining what an ``arm'' is, leading us to prefer
the term ``arm segments''.

\end{abstract}

\maketitle

\section{Introduction}
\label{sec:Intro}
\vspace{-3mm}

The Hubble Ultra Deep Field represents about 1/13,000,000 of the celestial
sphere and contains about 10,000 galaxies, suggesting
the entire sky contains upwards of $10^{11}$ galaxies.
Gaining quantitative structural information for this number of
galaxies will require automated methods.
For spiral galaxies, existing methods for visual classification
\cite{Hubble1936,DeVaucouleurs1959} are either subjective or
non-quantitative, while currently available semi-automated methods
\cite{DeSouza2004,Simard1998,Peng2002a,Peng2010,BenDavis2012,Ma2001,Au2006,Ripley1990,Shamir2011}.
are either too simplistic or require significant human input.

\begin{figure*}[t]
\plotsix{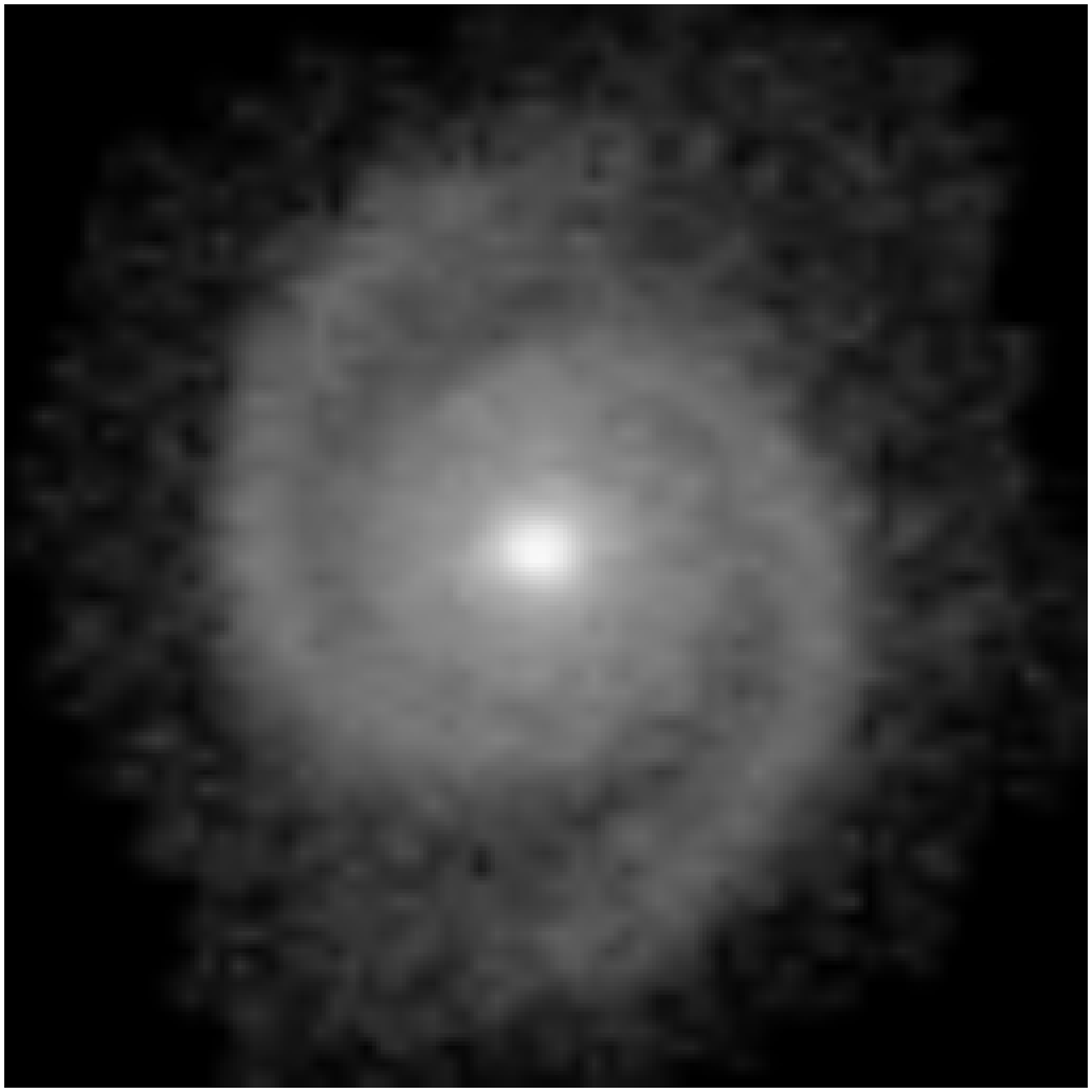}{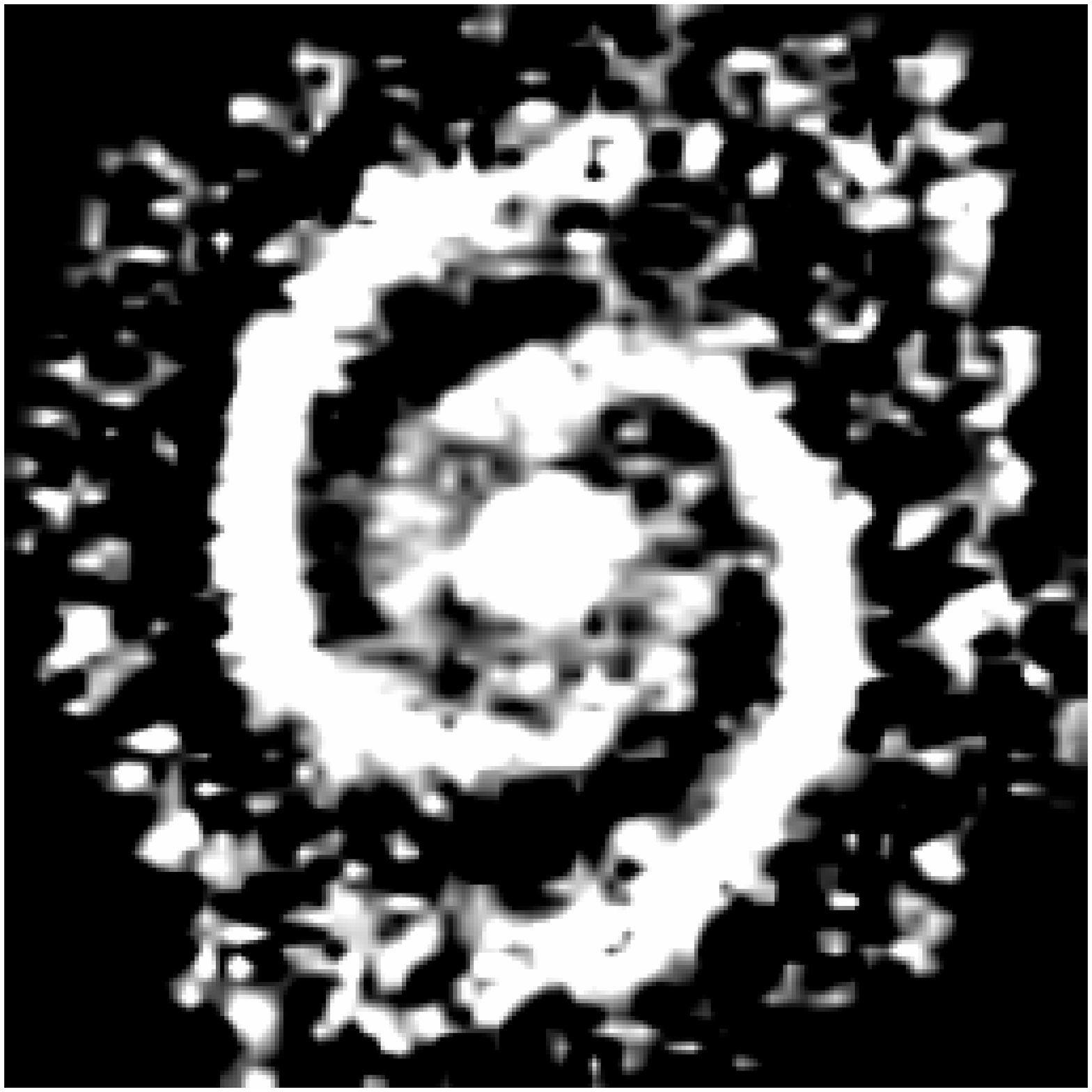}{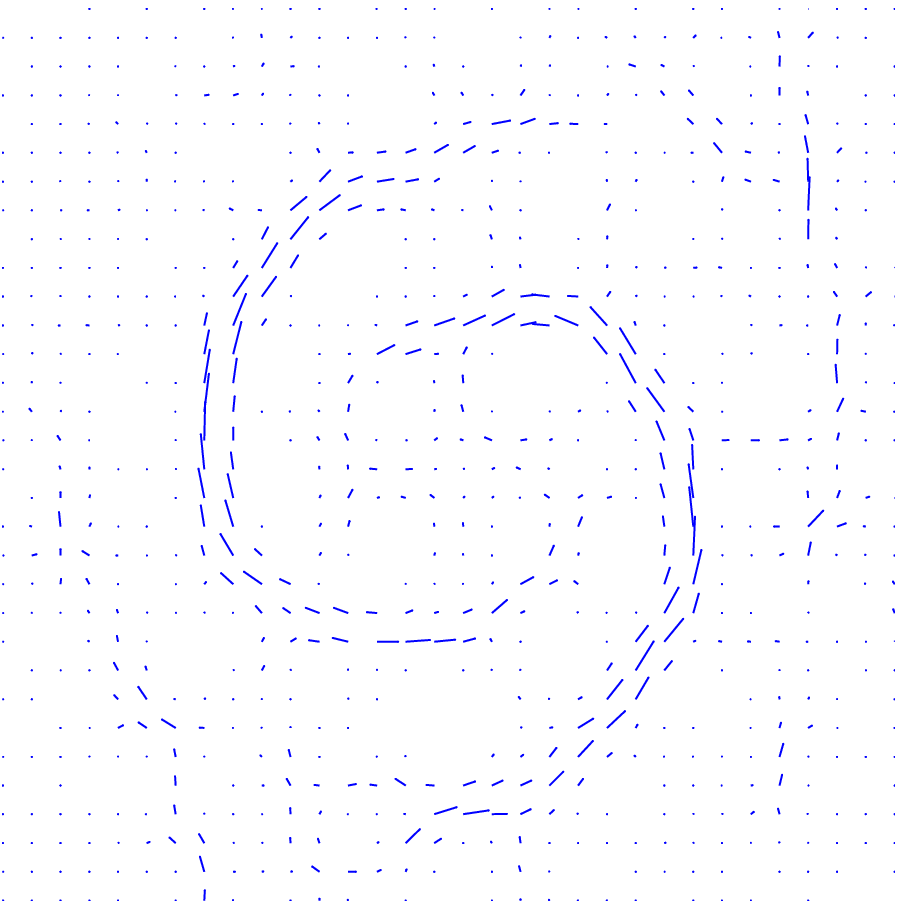}{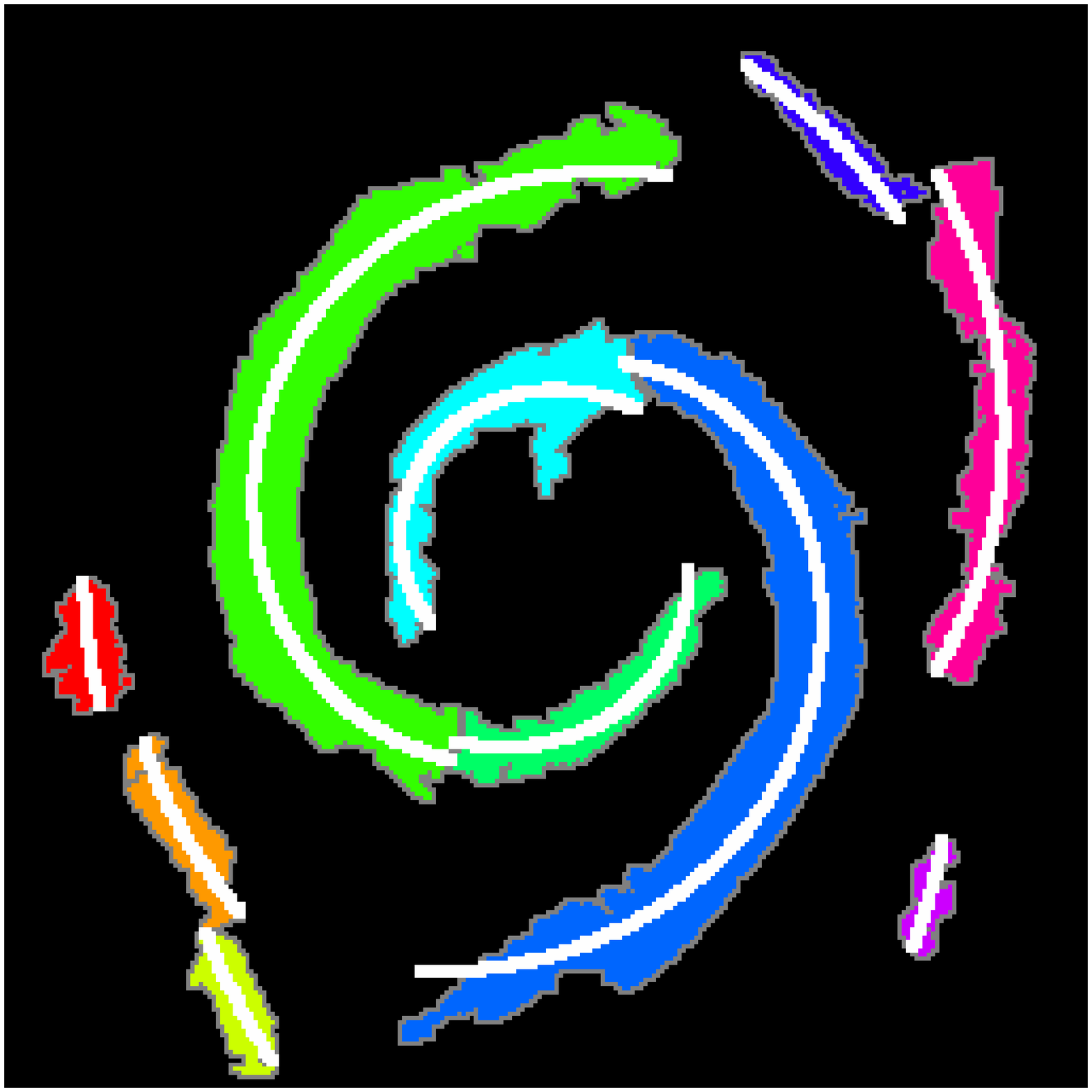}{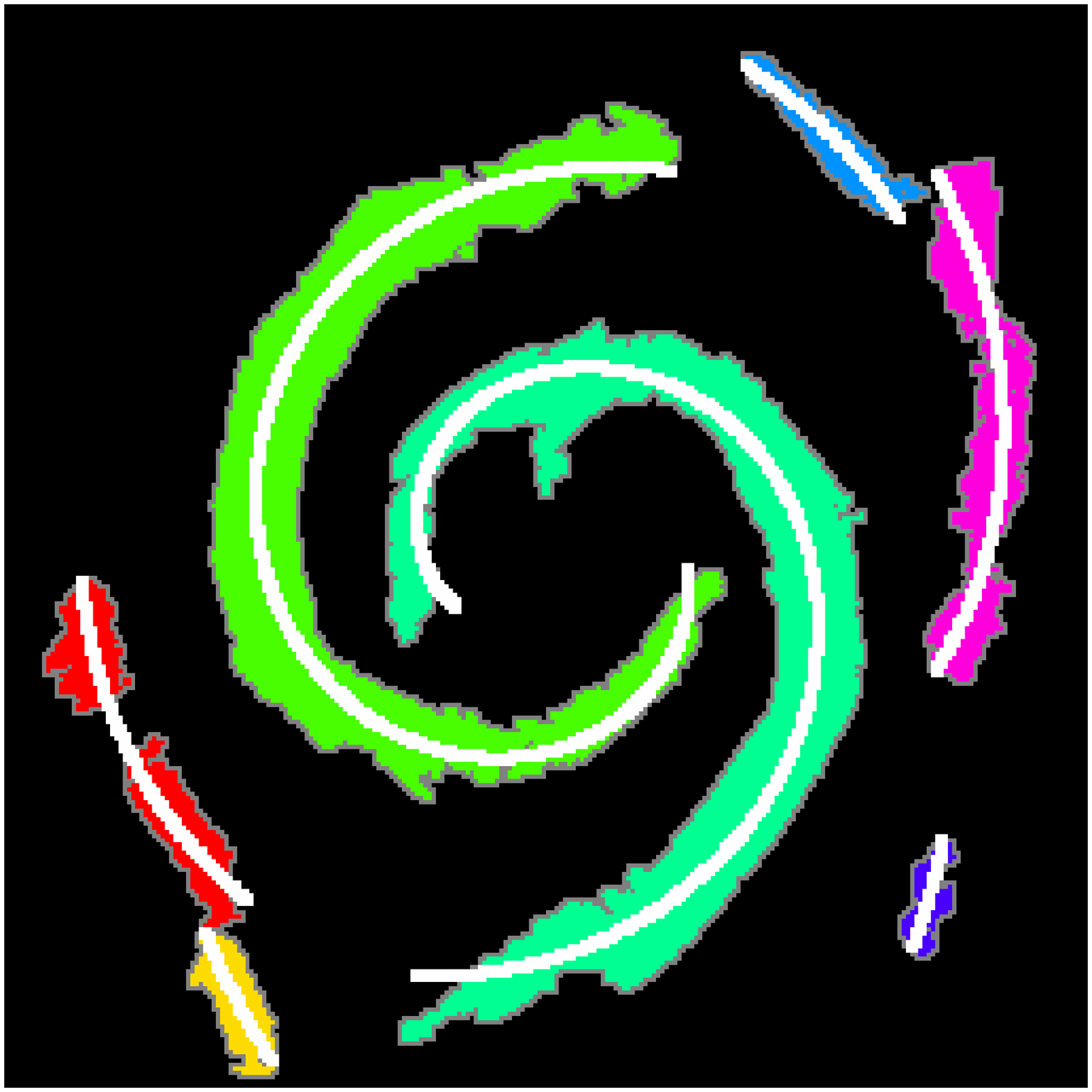}{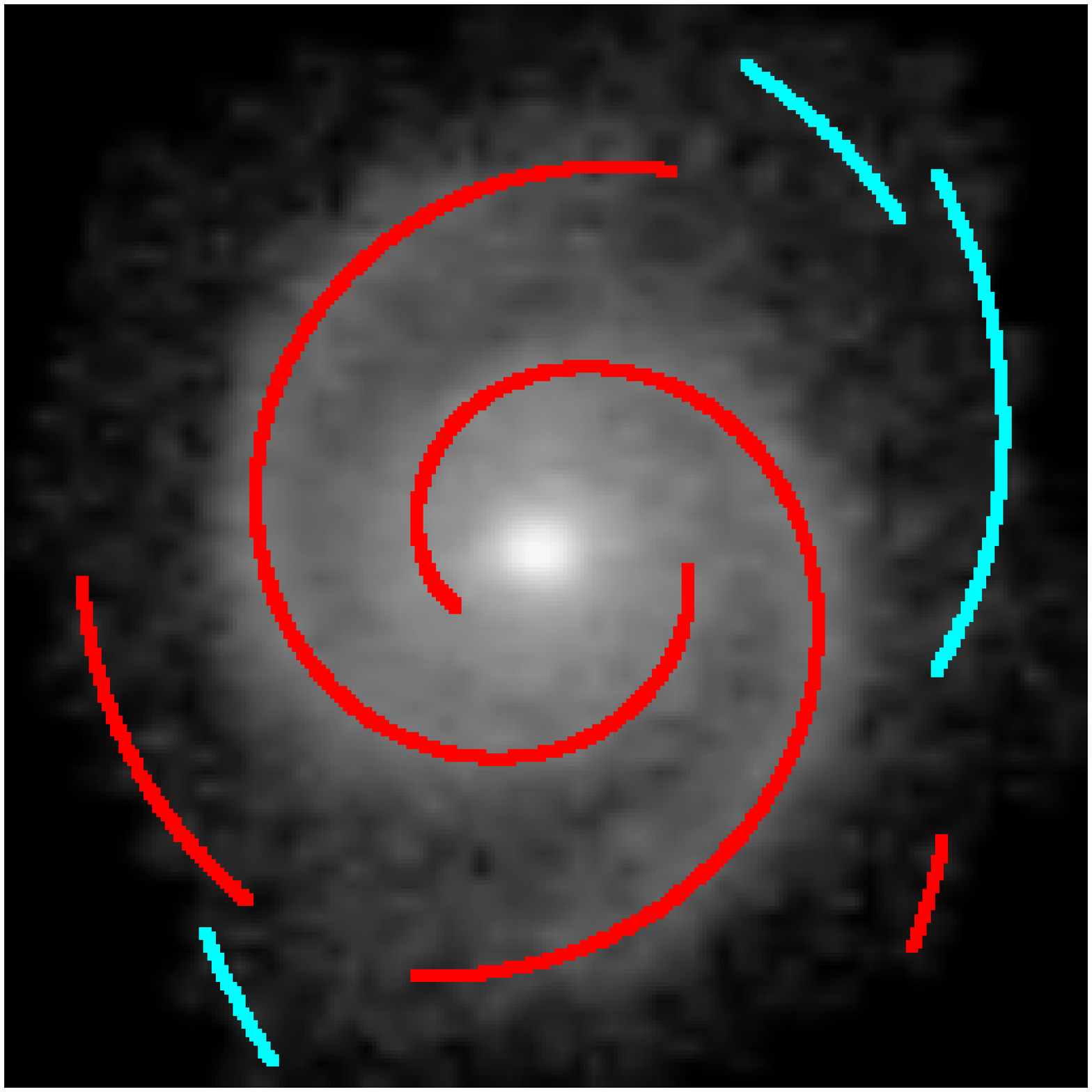}
$\begin{array}{cccccc}
\;\;\;\;\;\;\;\;\;\;\;\;\;\,  ({\bf a}) \;\;\;\;\;\;\;\;\;\;\;\;
&\;\;\;\;\;\;\;\;\;\;\;\, ({\bf b}) \;\;\;\;\;\;\;\;\;\;\;\;\;\;\;\;
&\;\;\;\;\;\;\;\;\;\;\; ({\bf c}) \;\;\;\;\;\;\;\;\;\;\;\;\;\;\;\;
&\;\;\;\;\;\;\;\;\;\;\;\; ({\bf d}) \;\;\;\;\;\;\;\;\;\;\;\;\;\;\;
&\;\;\;\;\;\;\;\;\;\; ({\bf e}) \;\;\;\;\;\;\;\;\;\;\;\;\;\;\;
&\;\;\;\;\;\;\;\;\;\;\; ({\bf f}) \;\;\;\;\;\;\;\;\;\;\;\;\;\;
\end{array}$
\caption{Steps in describing a spiral galaxy image.
{\bf a)} The centered and de-projected image.
{\bf b)} Contrast-enhanced image.
{\bf c)} Orientation field (at reduced resolution for pedagogical reasons).
{\bf d)} Initial arm segments found via Hierarchical Agglomerative Clustering (HAC) of nearby pixels with similar orientations and consistent logarithmic spiral shape, and the associated logarithmic spiral arcs fitted to these clusters.
{\bf e)} Final pixel clusters (and associated arcs) found by merging compatible arcs.
{\bf f)} Final arcs superimposed on image (a).  Red arcs wind S-wise, cyan Z-wise.
}
\label{fig:parse}
\end{figure*}

\section{Methods}
\vspace{-3mm}

Our method is entirely automated, and is described in detail elsewhere \cite{DavisHayesCVPR2012}.
Starting with an approximately
centered image of a galaxy, it uses an iterative 2D Gaussian fit to
find, precisely center, and size the image.

We make the simplifying assumption that the disk of the galaxy
would be circular if viewed face-on,
and then de-project it to produce a face-on view of the galaxy
(Figure \ref{fig:parse}a).
After applying a contrast-enhancement filter (Figure \ref{fig:parse}b),
an orientation filter \cite{Au2006} is used to assign an
orientation (strength and direction) to each pixel in the image based on the pixels around it
(Figure \ref{fig:parse}c).  Pixels are then hierarchically clustered
into regions with locally similar orientation and consistent logarithmic
spiral shape (Figure \ref{fig:parse}d).
{\em Note that brightness plays no explicit role in this clustering, although
it plays an implicit role through the creation of the orientation field.}
Each cluster is associated with a logarithmic spiral arc determined
by a least-squares fit, which can be brightness-weighted
(also Figure \ref{fig:parse}d).
Sometimes, the requirement for consistent logarithmic spiral structure
blocks merges of clusters that, in retrospect, "should" have been merged.
Later, however, as the clusters grow into their final shape, the arc
fits may become more compatible than they first appeared. Thus, a second
stage of merging is performed,
based primarily upon compatible spiral arc parameters
(Figure \ref{fig:parse}e).
Figure \ref{fig:parse}f depicts the resulting arcs overlayed on the original de-projected image.

The resulting arcs and arm segments are independent of each other,
do not need to conform to any symmetry criterion or be attached to a
bar or the bulge, and do not need to all wind in the same direction.  So
far as we are aware, this list of arm segments comprises the most detailed
description currently available for general spiral structure.
Given the list of pixels for each segment, an astronomer could
easily perform whatever measurement of that segment they may wish, such
as color, luminosity, brightness profile, etc.

\begin{figure*}[htb]
\centering
\plotsix{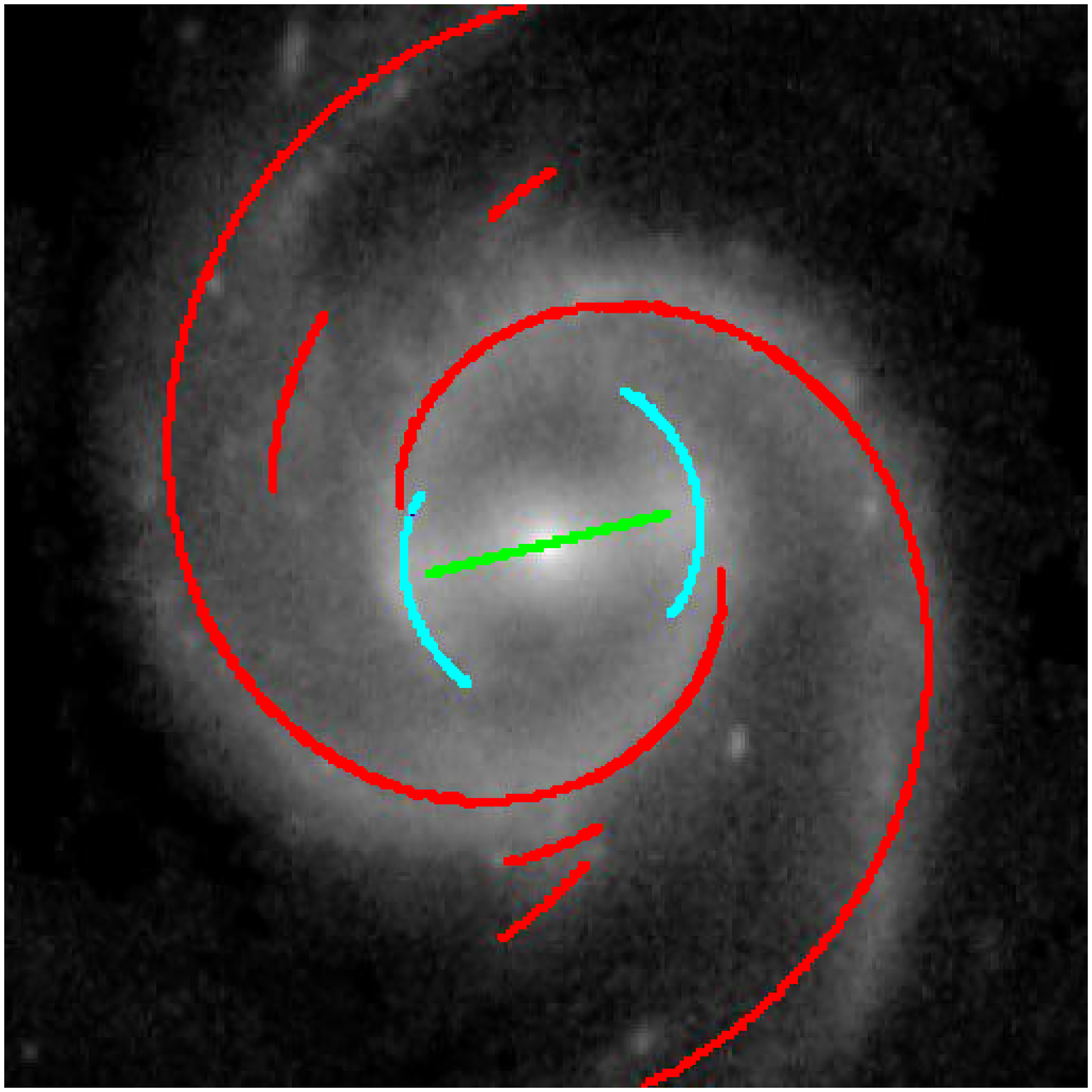}{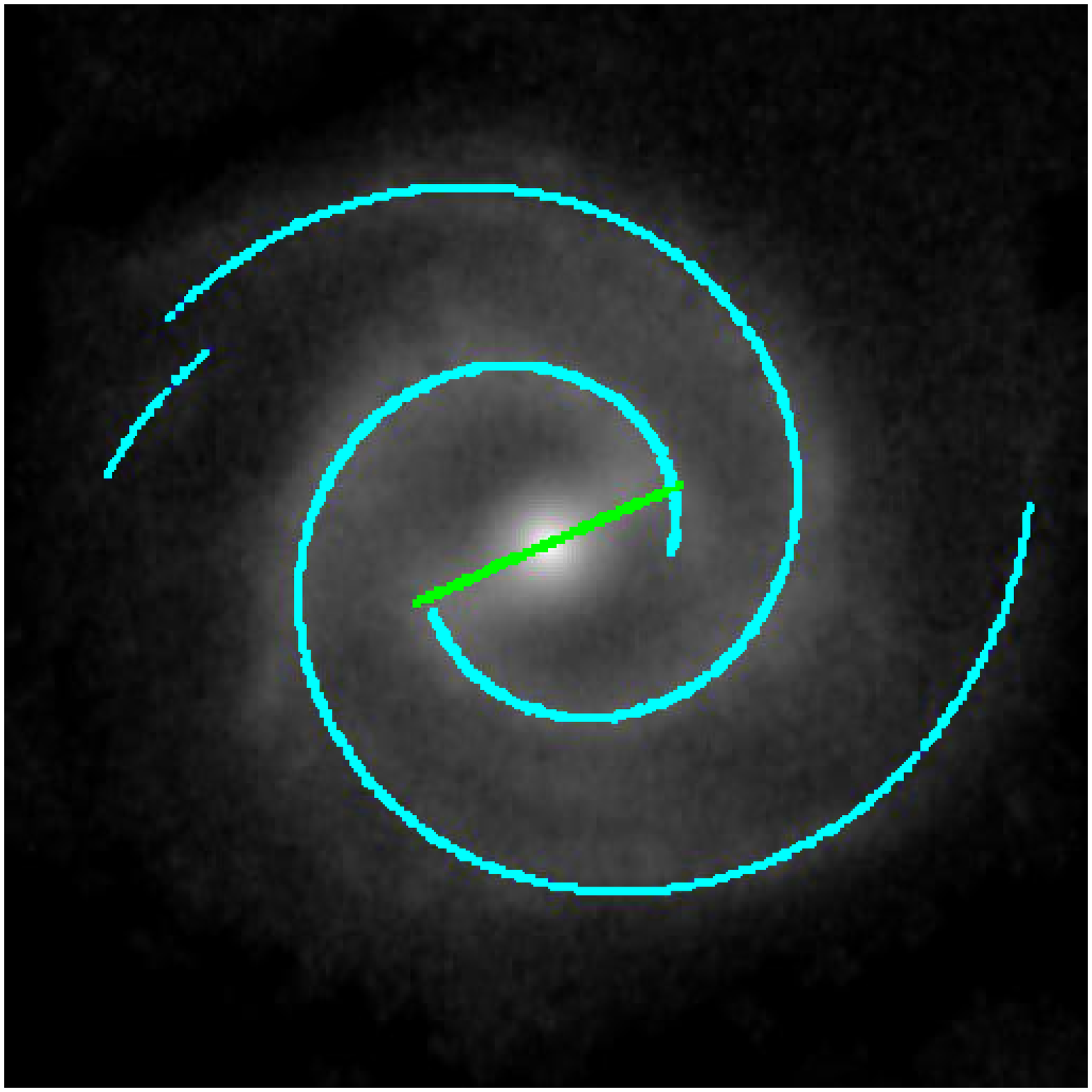}{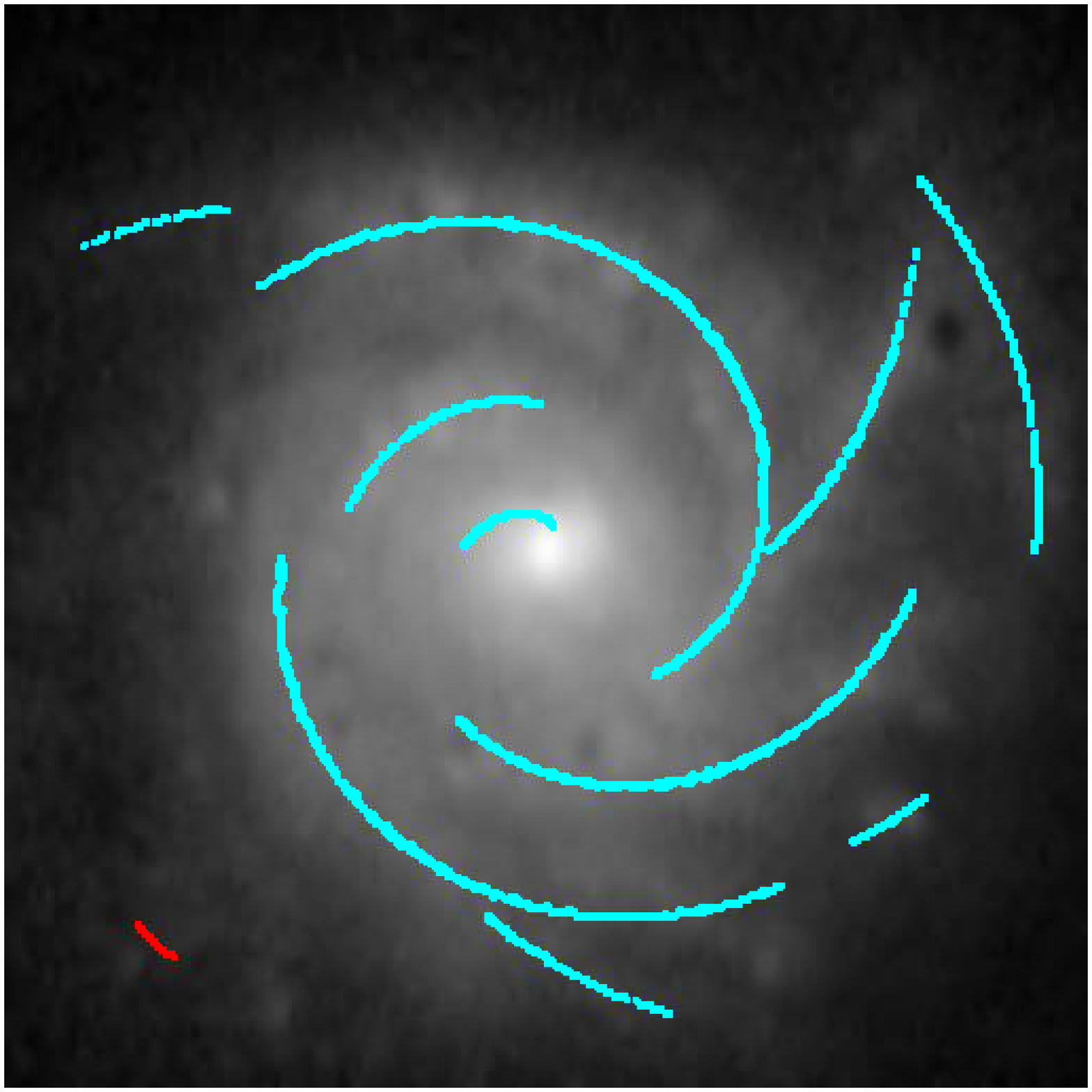}{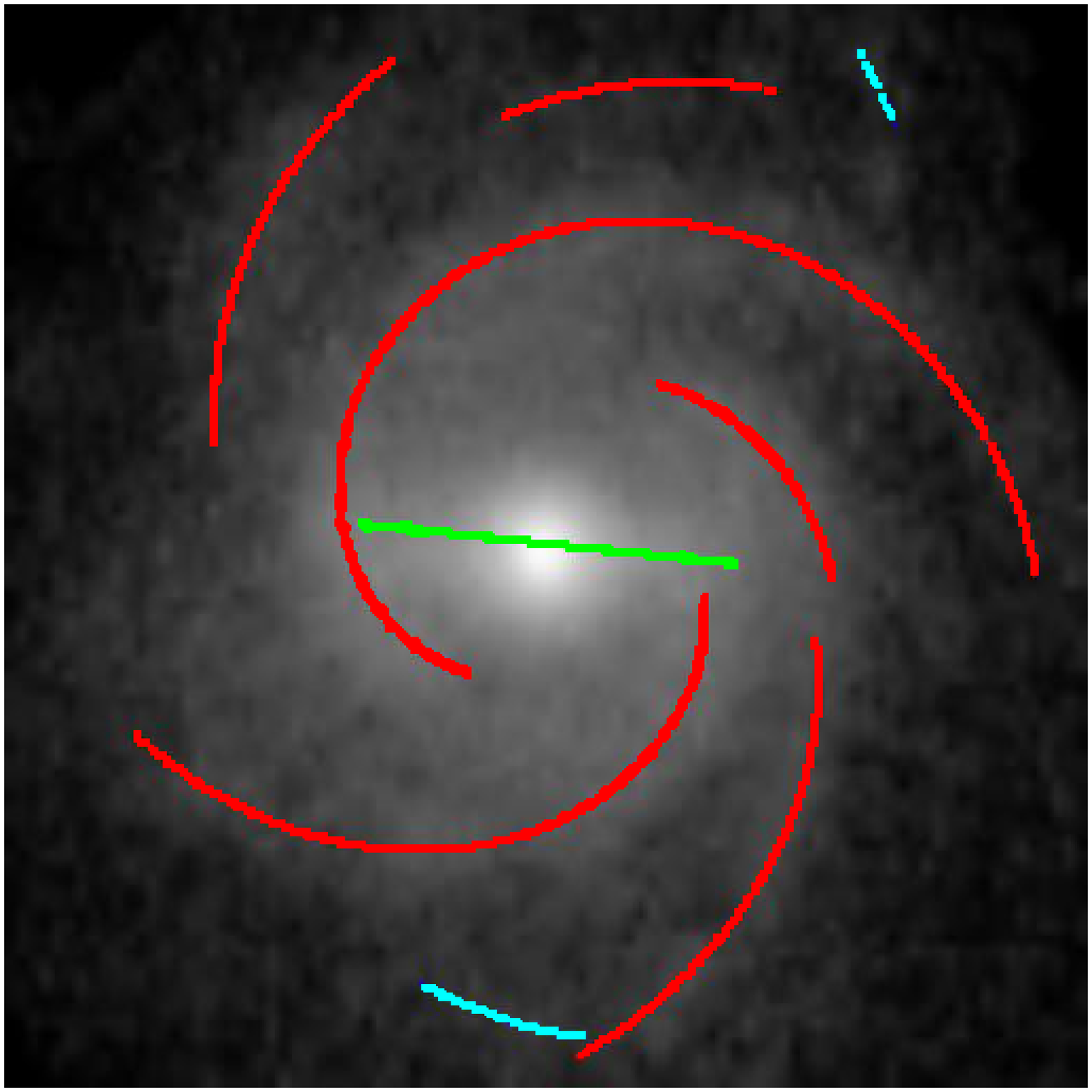}{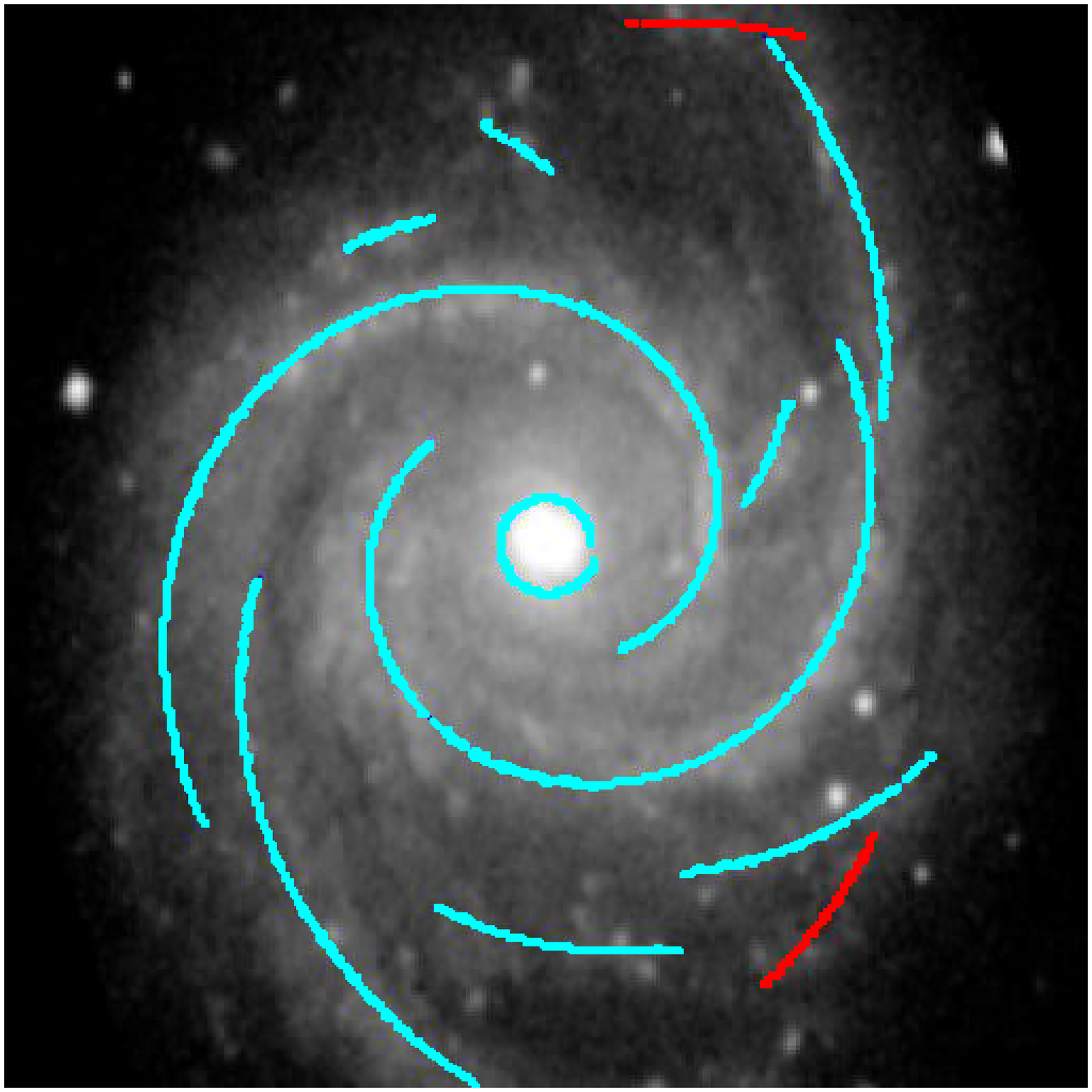}{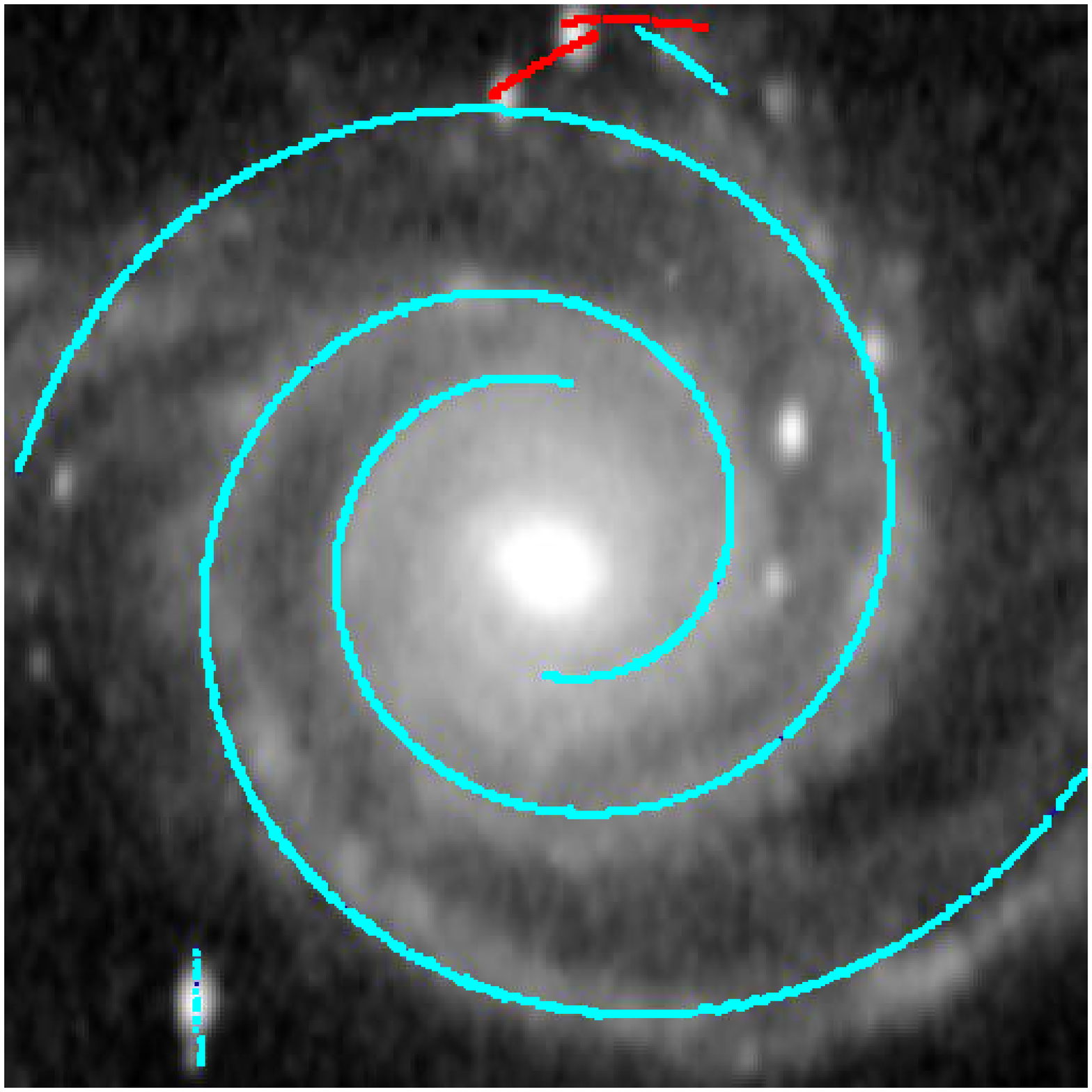}
\caption{Typical examples of how the algorithm performs on ``nice'' galaxies. Green represents a bar; red arms wind S-wise, cyan arms wind Z-wise.}
\label{fig:nice}
\end{figure*}

When run on a set of galaxies, our algorithm produces two tables: one table lists
every arc in every galaxy, while the second offers a per-galaxy summary.

\begin{figure*}[htb]
\plotfive{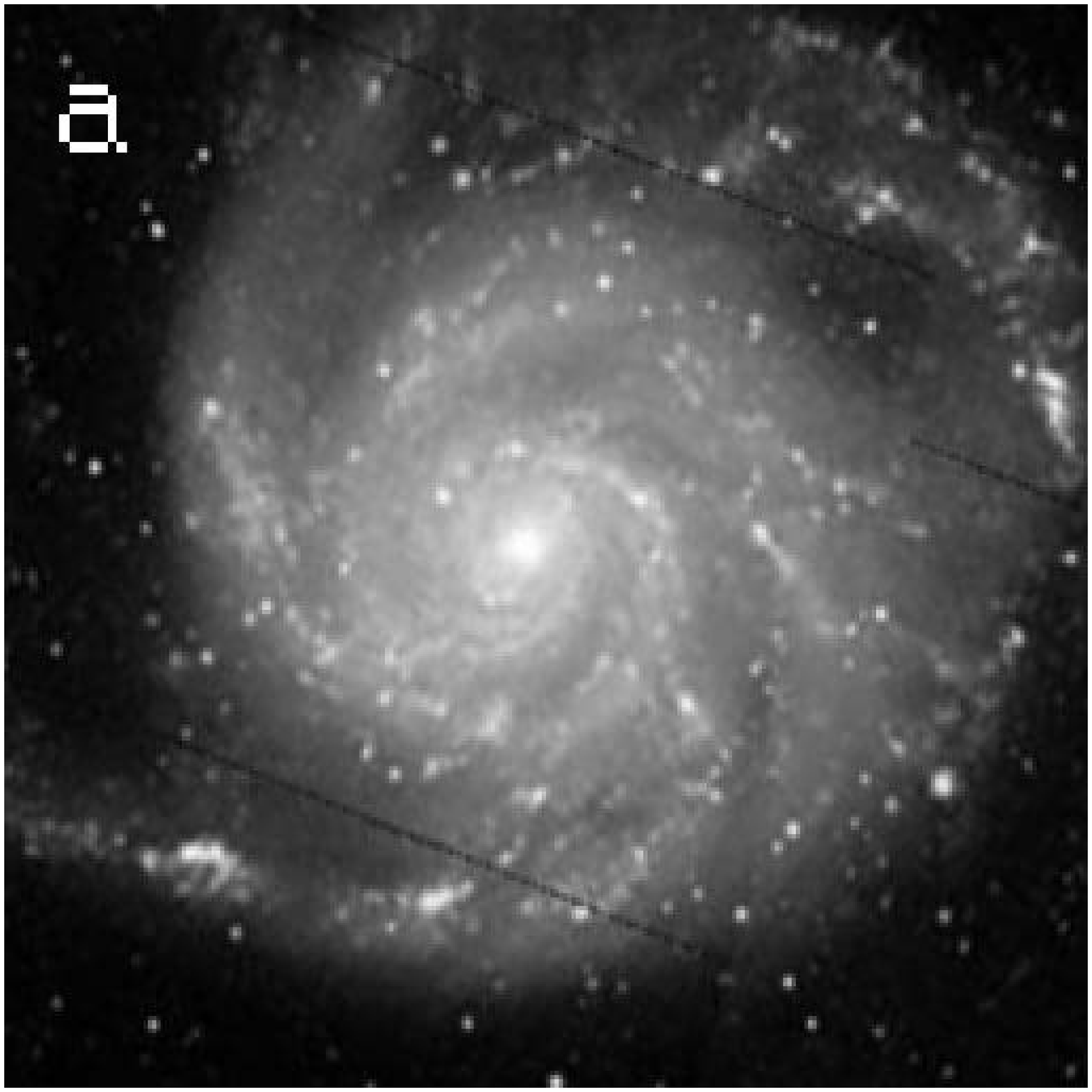}{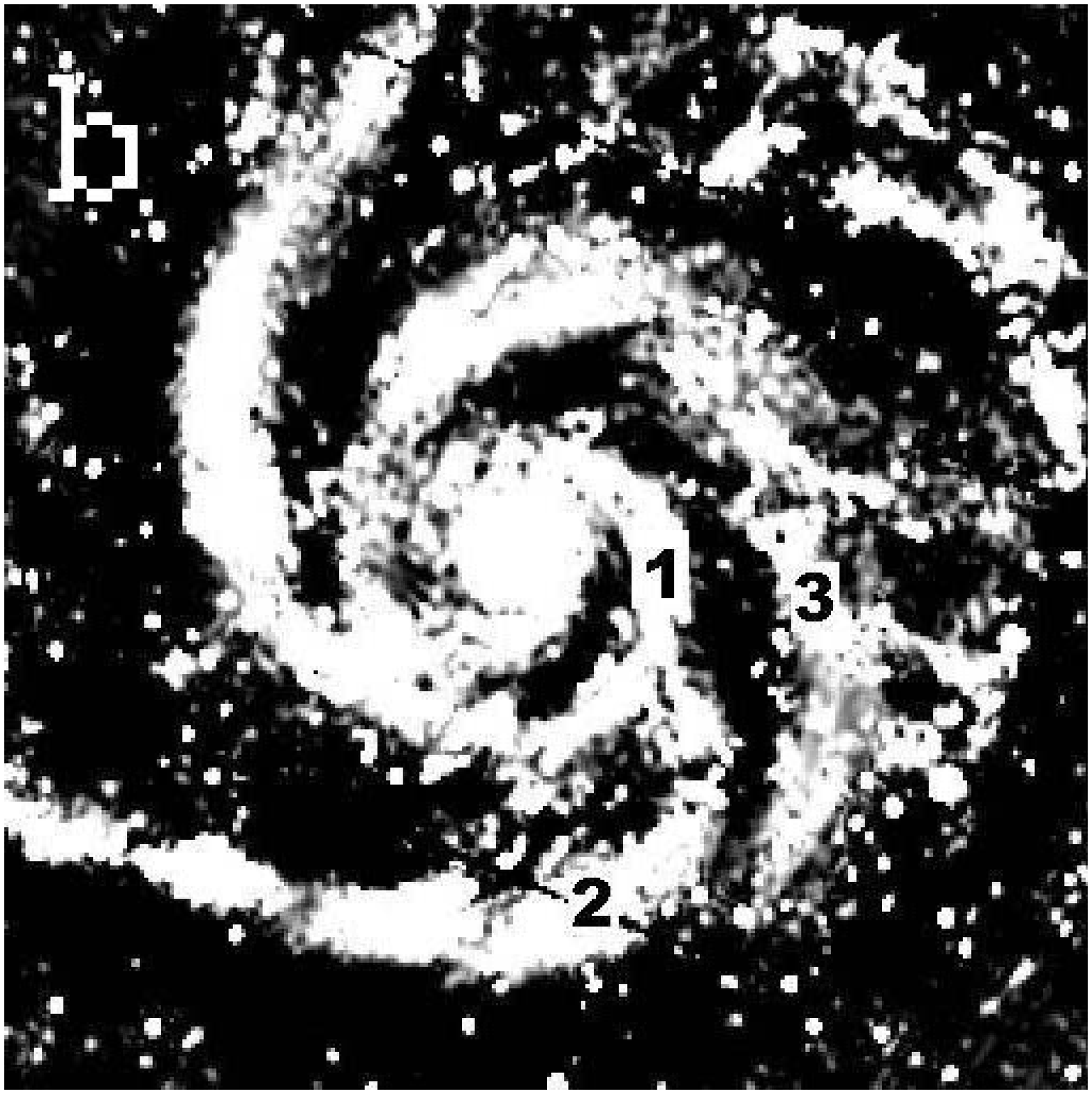}{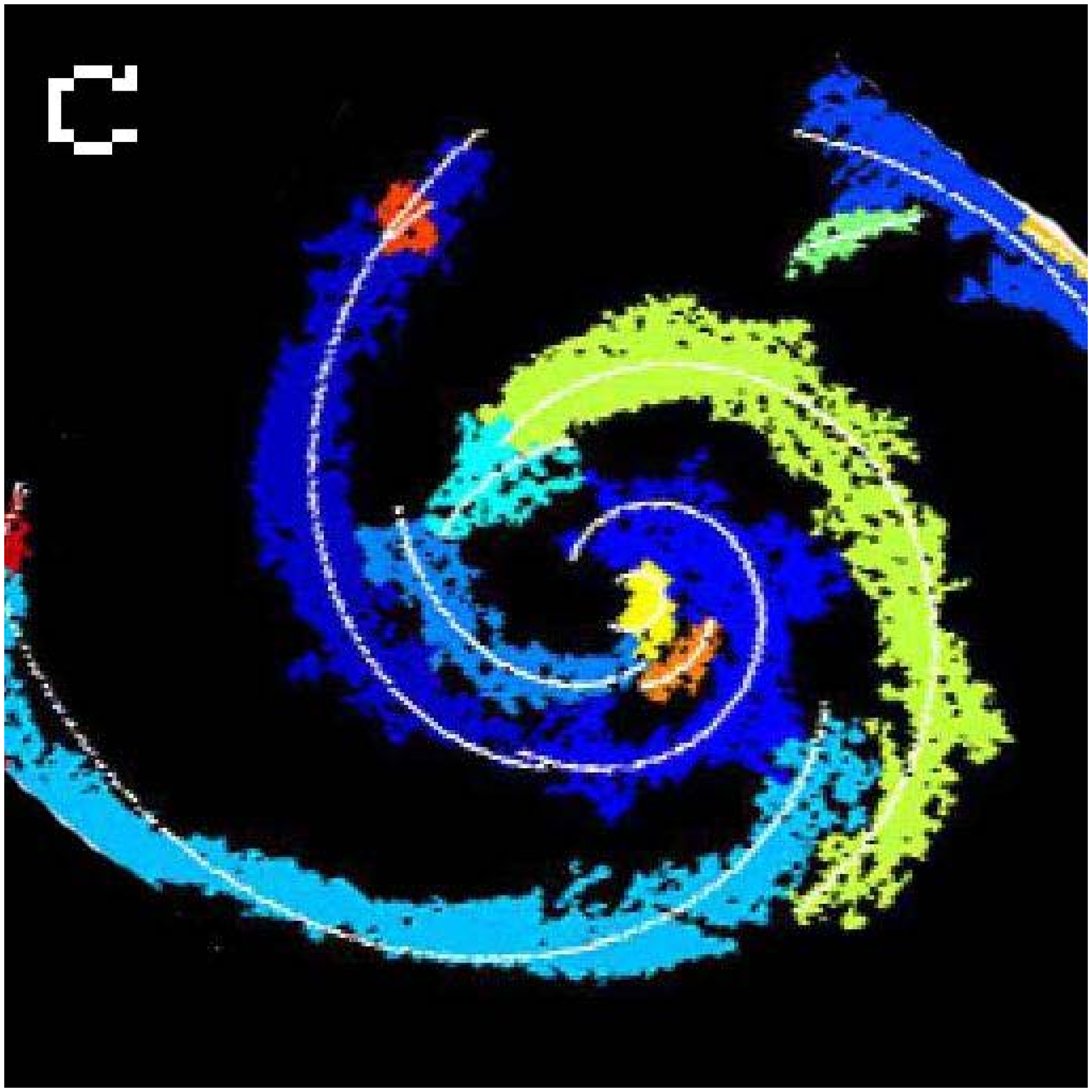}{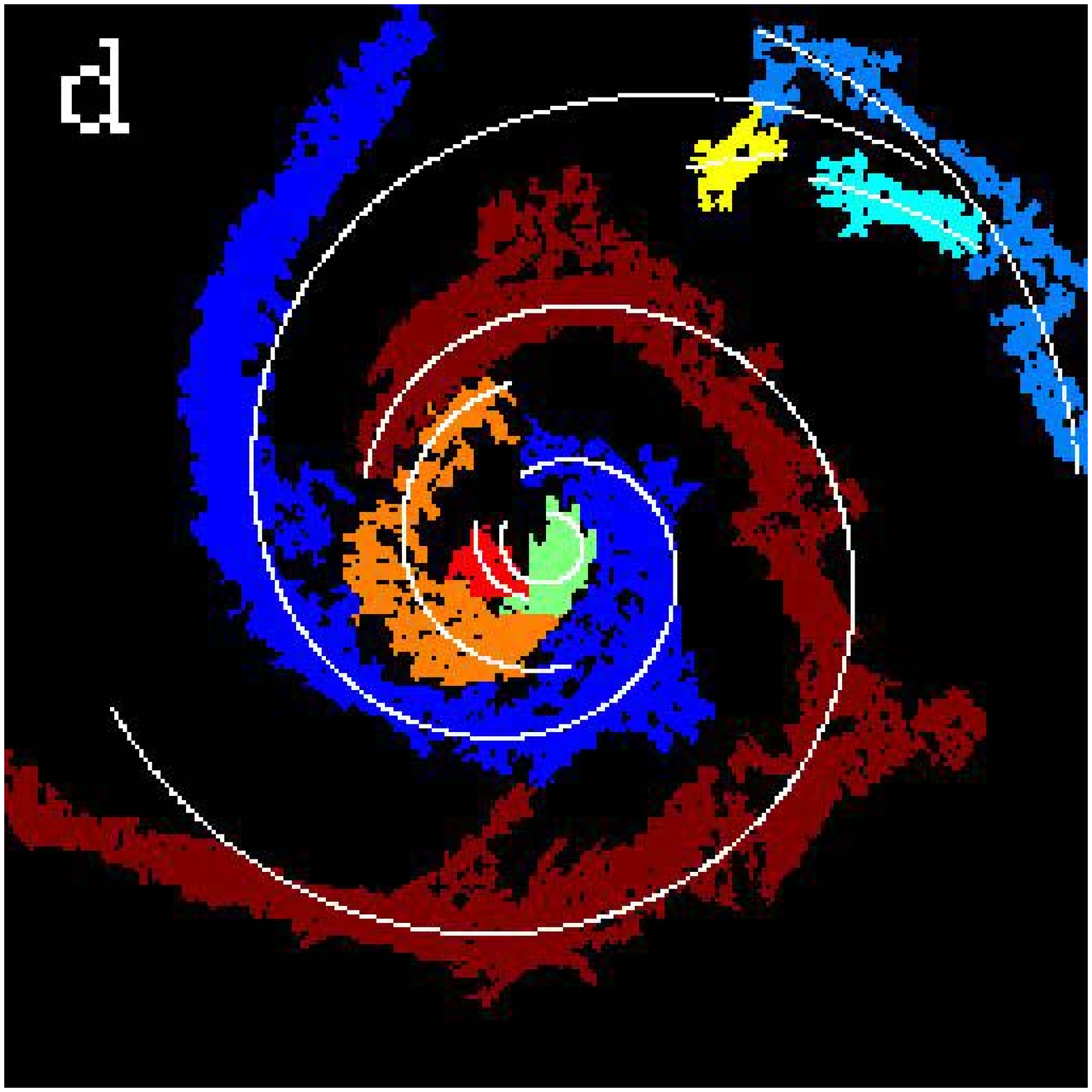}{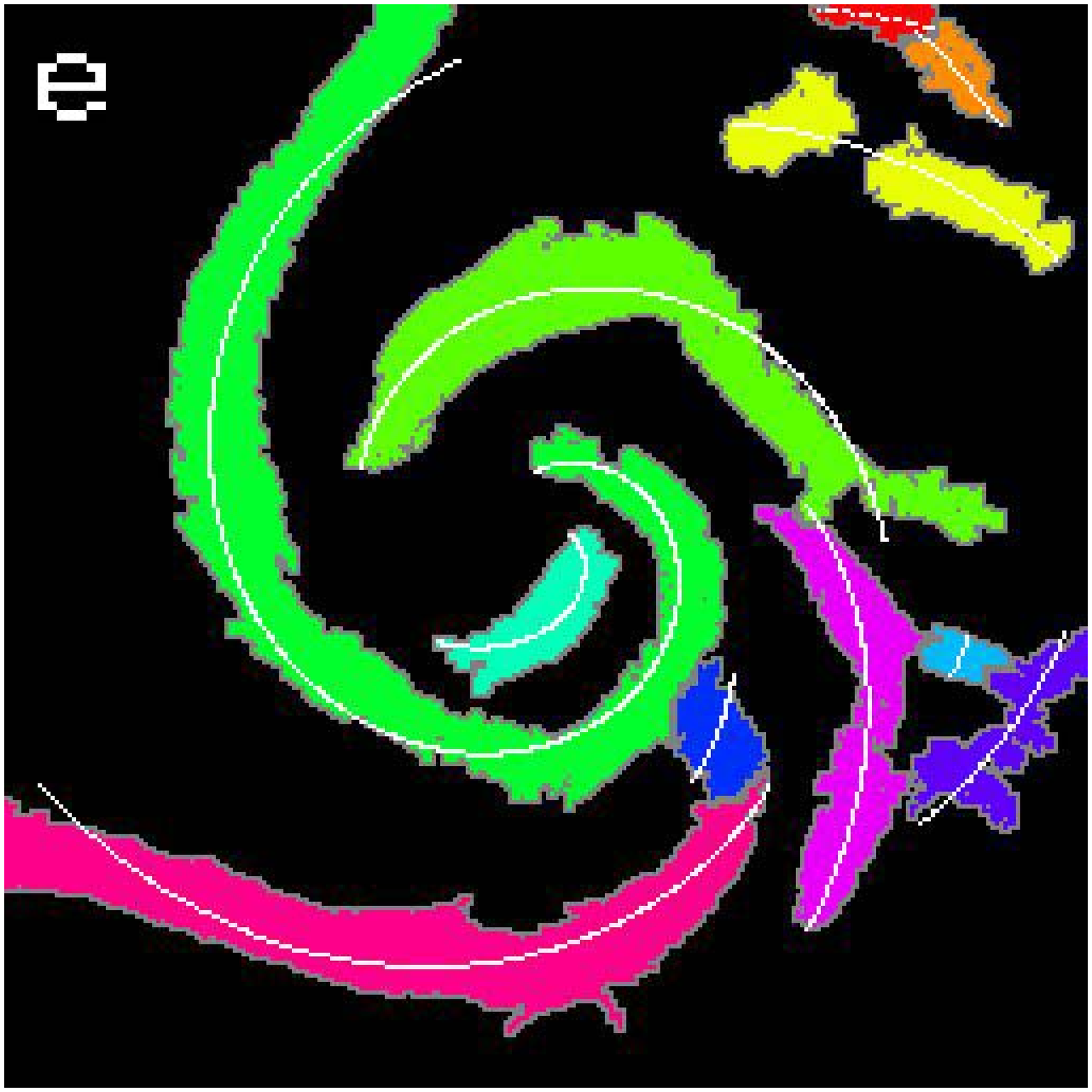}
\caption{Three different interpretations of the spiral structure in M101.
The colored images are from different versions of the code.  All comprise reasonable interpretations of
the structure.
{\bf (a)} original image
{\bf (b)} Contrast enhanced. We have labelled three arm segments; the joint between segments 1 and 2 may be what
a human would call a ``fork'', although our code never refers to forks.
{\bf (c)} An old version of the code, where the three segments happen to be separated (blue, cyan, and olive pixel clusters).
{\bf (d)} An intermediate version of the code, where segment 2 has been interpreted as a continuation of segment 3, jumping over the gap between them.  The single logarithmic spiral arc spanning the two arguably fits reasonably well,
suggesting perhaps that segments 2 and 3 are physically one arm with an obscuring dust lane, while the apparent
``fork'' of segment 2 from segment 1 is an optical illusion.
{\bf (e}) The most recent version of the code, in which the logarithmic spiral arc more
stingently fits (in the least-squares sense) each cluster of pixels.
}
\label{fig:WhatsAnArm}
\end{figure*}

\section{Results}
\vspace{-3mm}

Figure \ref{fig:nice} shows some typical ``nice'' examples.
In general, we find that if the image is clear and 
has a sufficient signal-to-noise ratio (about as good as needed by a human,
no more), then the algorithm does a
very good job of marking out the arms and determinining pitch angle.
However, we believe the strength of our algorithm is that it {\em also}
does well when the galaxy is less clean, less symmetrical, and
more fragmented.  Even when the image has such low resolution that little
structure is visible, we find good agreement with human determinations
of structure.

Observing the arcs superimposed on the images in Figures
\ref{fig:parse}, \ref{fig:nice} and \ref{fig:WhatsAnArm}, we make two observations.
First, that the logarithmic spiral arc is a very good mathematical description
of the curve of spiral arms (with the possible exception of Figure \ref{fig:WhatsAnArm}d,
which is a version of the code that is no longer in use and did not always properly handle
multi-revolution arc fits).
And second, that even though the fit is usually very good,
the pitch angle between different arms in the same galaxy can be quite different.
Even the two longest arms in the galaxy of Figure \ref{fig:parse} do not agree with each other:
the ``left'' long arm in Figure \ref{fig:parse}f has a pitch angle of 13.6
degrees while the ``right''
long one has a pitch angle of 24.4 degrees---a difference of almost 11 degrees.
Furthermore, it is clear from Figure \ref{fig:parse}e that both arms are fit quite well, with a width much
smaller than their length and the arcs remaining inside their cluster for their entire
length; this indicates that the error in these pitch angles is much less than
their difference.
Such a disagreement is not atypical, as Table \ref{tab:median_pa_diff} shows.
Recent work has suggested that pitch angle can also vary with radius, and this
variance correlates with other structural parameters of the galaxy \cite{SavchenkoReshetnikov2013}.

\begin{table}[h]
\footnotesize
\centering
\begin{tabular}{|r|rrrrrrr|}
\hline
minimum length &     0 &   50 &   100 &  150 &  200 &  250 &  300 \\
\hline
median difference & 14.5 & 14.3 &  10.7 &  7.5 &  5.6 &  3.5 &  2.6 \\
\hline
\end{tabular}
\caption{Median pitch angle difference (bottom row, in degrees) between
the two longest arms in a galaxy when both are at least as long (in
pixels) as the top row.  Images are scaled to 256x256 pixels.}
\label{tab:median_pa_diff}
\end{table}

Astronomers frequently refer to ``spiral arms'' in a galaxy, but to our knowledge
there does not exist a formal definition of what, exactly, a spiral arm is.
Figure \ref{fig:WhatsAnArm} illustrates why this may be the case: arm segments
can fork, split, have dust lanes, and have ``kinks'' in them where the pitch
angle changes even if they are visually contiguous.  As a result, in
this paper we prefer to use the term ``arm segments'', which we roughly define
as an (almost) contiguous arc-shaped region of light with locally consistent
orientation.
Figure \ref{fig:WhatsAnArm} also gives an informal indication of how the parsing
of an individual image can change depending upon minor changes to the algorithm.

Given a fixed version of the algorithm in which only its major parameters are changed,
it is important to determine that our results change in a reasonable way as a
function of the parameters.
Although the results for an individual galaxy may occasionally change
abruptly as algorithmic parameters change---especially if the galaxy is poorly
resolved---the average results across a large sample of
galaxies should change smoothly.  We have performed a sensitivity analysis
and verified that, statistically, all output variables vary smoothly.  Furthermore,
both the mean and median pitch angle across galaxies changes by less than a degree
as algorithmic parameters are varied by a factor of 2 in either direction.
This gives us confidence that the results of our algorithm are stable in
the sense that statistics across a large sample of galaxies are unlikely
to suffer a large or unexpected shift based on small changes to the parameters.

\begin{figure}[htb]
\centering
\includegraphics[scale=0.95]{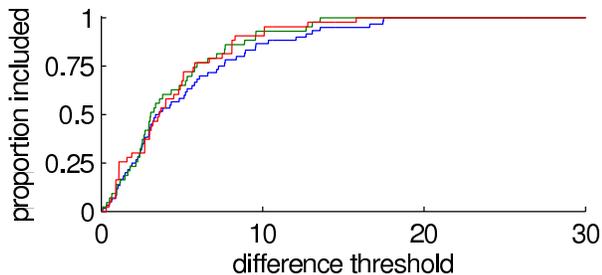}
\caption{Cumulative distribution of pitch angle discrepancies between the two arcs measured for each galaxy from \protect\cite{Ma2001}, in red, and between our measurements and the measurements in \protect\cite{Ma2001} (blue for the full comparison set, and green for the subset where two arc measurements are available from \protect\cite{Ma2001}).
Since the two curves are very similar, it is likely that much of the scatter 
between our results and the results of other authors
arises from genuine within-galaxy arm variation rather
than between-method measurement variation.
}
\label{fig:JunMaDiffVsConsistency}
\end{figure}

So far as we are aware, the current work is the first large-scale, automated,
quantitative survey of detailed general spiral structure in galaxies.
There exist several small-scale quantitative surveys (\eg \cite{Ma2001,BenDavis2012}),
and two large-scale, qualitative, human-based surveys: one citizen science
project (Galaxy Zoo \cite{Lintott2008,Lintott2010}), and one professional
survey \cite{Nair2010a}.

We compare our algorithm to two small-scale pitch angle surveys.
In the first Ma \cite{Ma2001} manually measured the pitch angles of either one
or two arms in each of a small sample of galaxy images.
We have downloaded and run our algorithm on the images used in that paper.
The second (a group from the University of Arkansas) uses a Fourier Analysis to extract the dominant pitch
angle \cite{Seigar2006, BenDavis2012}.
In both cases, we always agree in chirality, although there is significant scatter,
sometimes up to 10--20 degrees, when it comes to pitch angle agreement.
Although this scatter may seem to be a concern, recall
from Table \ref{tab:median_pa_diff} that the pitch angle can vary quite significantly
between arms in a single spiral galaxy. Since Ma \cite{Ma2001} measured only one or
two arms in each galaxy, and since the Arkansas group measure only one dominant
pitch angle, perhaps the scatter could be explained by inter-arm differences
in each galaxy, rather than differences as a function of method.  To test this
hypothesis, Figure \ref{fig:JunMaDiffVsConsistency} plots the cumulative distribution
of pitch angle discrepancies between the two arcs in one galaxy (when available) from
Ma \cite{Ma2001}, {\it vs.} discrepancies between Ma's method and our method.

\begin{table}[t]
\footnotesize
\begin{center}
\begin{tabular}{|l|c c c c | c c c c |}
\hline
\textbf{min discernibility rate} & 0  & 60 & 90 & 100& 0  & 60 & 90 & 100 \\ \hline
\textbf{require 2 longest agree} & N  & N  &  N &  N & Y  & Y  & Y  & Y \\ \hline \hline
\textbf{inclusion rate}           &99.3&95.1&52.2&12.4&67.0&64.7&39.0&10.0\\\hline\hline
\textbf{per-arc majority vote}            &75.4&75.9&79.9&82.5&79.5&79.8&83.1&84.8\\\hline
\textbf{longest arc alone}         &84.9&85.3&89.4&92.4&95.3&95.7&97.8&98.4\\\hline
\textbf{length-weighted vote}       &89.4&89.9&93.5&95.7&94.9&95.3&97.5&98.4\\\hline
\end{tabular}
\end{center}
\caption{Winding-direction agreement with human classifications from Galaxy Zoo 1. Row 1: the minimum proportion of human votes that the dominant winding direction must receive (out of the two known-direction and four unknown-direction categories).  Row 2: do we demand our two longest arcs agree in chirality? Row 3 gives the proportion of the 29,250 galaxies included under these criteria. Rows 4, 5, and 6 give agreement rates between Galaxy Zoo and three methods of determining winding direction from our output.}
\label{tab:GZChiralityAgreement}
\end{table}

Galaxy Zoo \cite{Lintott2008,Lintott2010} is a citizen science project in which
approximately 250,000 human volunteers classify images of galaxies over the web after
some rudimentary training.  The median number of people who viewed each image was about 40,
so that some measure of certainty can be obtained from multiple viewings of each image.
Galaxy Zoo 1 (GZ1, \cite{Lintott2010}) presented people with an image of a real galaxy
along with 6 cartoon galaxies, and asked them to choose which of them
most resembled the real galaxy. Across spiral galaxies with observable structure,
the only comparison we can make with GZ1 is chirality (S-wise vs.
Z-wise).  In difficult cases some humans may choose ``spiral'' while others do not,
so we compare our results on chirality against what we call the ``discernability'' of a galaxy:
the maximum fraction of agreeing humans that saw spiral structure, which we define for
a particular galaxy as
$\frac{\mbox{max(S-wise votes, Z-wise votes)}}{\mbox{total number of votes}}$.

Table \ref{tab:GZChiralityAgreement} compares our chirality measurement against those
of Galaxy Zoo 1 humans as a function of both human discernability, and several measures of
our output.  We include 29,250 galaxies with clearly observable
structure, chosen by the director of the GZ project (Steven Bamford, personal communication).
We see that both the ``Longest arc alone'' and
``Length-weighted vote'' agree with the humans at least 85\% of the time, and as much
as 98.4\% of the time, across all values of human discernability.

\begin{figure}
\centering
\includegraphics[scale=0.9]{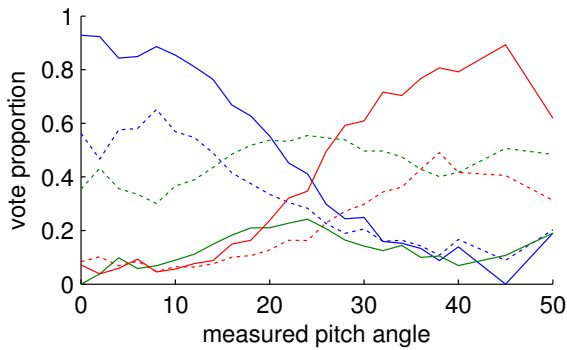}
\caption{Proportion of galaxies receiving a majority vote for Tight
(blue lines), Medium (green lines) or Loose (red lines) as a function of
our measured pitch angle. Pitch angles are binned with width 2 degrees
between 0 and 40, 5 degrees between 40 and 50, and one bin beyond 50
(due to low sample size). The dashed lines represent all images tested
from Galaxy Zoo 2; the solid lines represent the images within the top
quartile of human agreement.}
\label{fig:GZ2TightnessComparison}
\end{figure}

We also compare pitch angle (arm tightness) measurements, starting with Galaxy Zoo 2. Here, where human classifiers indicated that there was ``any sign of a spiral arm pattern'' \cite{Masters2011}, they were asked whether the spiral arms were tight, medium, or loose.
Figure \ref{fig:GZ2TightnessComparison} shows the relationship between our measured pitch angle (arclength weighted vote)
and the proportion of galaxies receiving a majority human vote for Tight, Medium, or Loose.
As can be seen from the dashed lines in Figure \ref{fig:GZ2TightnessComparison}, galaxies where we measure a low pitch angle usually have majority human votes for Tight, while most of the remaining galaxies in this range had majority votes for Medium. As our measured pitch angle increases, we see progressively fewer galaxies classified as Tight, and more galaxies classified as Loose. Designations as ``Medium'' are pervasive throughout, while ``Loose'' classifications are less frequent than ``Tight'' classifications. This reflects the classification distribution of the image set as a whole.
The solid lines in Figure \ref{fig:GZ2TightnessComparison} represent the top agreement quartile (lowest Shannon entropy quartile), and the association between our tightness measure and human classifications is even more pronounced.

Longo \cite{Longo2011} also provided a chirality measurement for a survey of galaxies,
in which each galaxy was viewed only once by one of five student volunteers.  The
students were told to choose a chirality only if it was clear.
We find similar agreement as was seen against Galaxy Zoo.

\section{Discussion and future work}
\vspace{-3mm}

We have run our code on every item in the Sloan Digital Sky Survey (SDSS)
that is classified as ``galaxy''.  Unfortunately, SDSS does not distinguish
between spiral and non-spiral galaxies.  We are currently working on a machine
learning algorithm that uses the output of our code to distinguish between
spiral and non-spiral galaxies.  Preliminary results are encouraging (agreeing
approximately 90--95\% with Galaxy Zoo humans), and will be presented in a future
paper.  Once we can separate out spiral galaxies, further studies will be performed
concerning how spiral structure correlates with other variables such as color,
redshift, local environment, etc.

Code and data are available from WBH upon request.

\vspace*{-3mm}\section*{Acknowledgments}
\label{sec:Acknowledgments}
\vspace{-3mm}
We thank Steven Bamford for helpful insights, image
sample selection, and pre-publication access to Galaxy Zoo
2 classifications; Charless Fowlkes, Deva Ramanan, Eric Mjolsness,
and Aaron Barth for helpful discussions; and the Arkansas
Galaxy Evolution Survey (AGES) Collaboration for pitch
angle measurements and discussions. Comparisons were
also made possible due to image data from CGS as well
as from SDSS and POSS II. Fellowship and travel support
was provided by an ICS Dean's Fellowship at UC
Irvine (for DD); the Oxford Centre for Collaborative Applied
Mathematics; Steven Bamford and the MegaMorph
project; and the AGES Collaboration (through NASA Grant
NNX08AW03A).


{\small
\bibliographystyle{chicago}

}

\end{document}